\documentclass[10pt,twocolumn,preprintnumbers,superscriptaddress,nofootinbib,aps,prl]{revtex4}

\usepackage{graphicx}
\usepackage{amsmath}
\usepackage{srcltx}
\usepackage[utf8]{inputenc}
\usepackage[colorlinks]{hyperref}
\usepackage{times}
\usepackage{xcolor}
\usepackage{amssymb}
\newcommand{\vev}[1]{\left\langle {#1} \right\rangle}
\newcommand{\lsim}{\lesssim}
\newcommand{\gsim}{\gtrsim}

\newcommand{\eq}[1]{Eq.~(\ref{#1})}
\usepackage[normalem]{ulem}

\newcommand{\ord}[1]{\mathcal{O}{\left(#1 \right)}}
\newcommand{\beq}{\begin{equation}}
\newcommand{\eeq}{\end{equation}}
\newcommand{\bea}{\begin{eqnarray}}
\newcommand{\eea}{\end{eqnarray}}
\newcommand{\eps}{\varepsilon}

\newcommand{\mbX}{\bar m_X}

\newcommand{\vtwo}[1]{\textcolor{black}{#1}}

\newcommand{\appropto}{\mathrel{\vcenter{
  \offinterlineskip\halign{\hfil$##$\cr
    \propto\cr\noalign{\kern2pt}\sim\cr\noalign{\kern-2pt}}}}}

\begin{document}

\pagestyle{plain}

\title{Archimedean Lever Leptogenesis
}

\author{Djuna Croon}
\email{djuna.l.croon@durham.ac.uk}
\affiliation{Department of Physics, Durham University, Durham DH1 3LE, UK}
\affiliation{Institute for Particle Physics Phenomenology, Durham University, Durham DH1 3LE, UK}

\author{Hooman Davoudiasl}
\email{hooman@bnl.gov}
\affiliation{High Energy Theory Group, Physics Department, Brookhaven National Laboratory, 
Upton, NY 11973, USA}

\author{Rachel Houtz}
\email{rachel.houtz@durham.ac.uk}
\affiliation{Department of Physics, Durham University, Durham DH1 3LE, UK}
\affiliation{Institute for Particle Physics Phenomenology, Durham University, Durham DH1 3LE, UK}


\begin{abstract}

We propose that weak scale  leptogenesis via $\sim 10$~TeV  scale right-handed neutrinos could be possible if their couplings had transitory larger values in the early Universe.  The requisite {\it lifted} parameters can be attained if a light scalar $\phi$ is displaced a {\it long} distance from its origin by the thermal population of fermions $X$ that become massive before electroweak symmetry breaking.  The fermion $X$ can be a viable dark matter candidate; for suitable choice of parameters, the light scalar itself can be dark matter through a misalignment mechanism.  We find that a two-component DM population made up of both $X$ and $\phi$ is a typical outcome in our framework.   
\end{abstract}

\preprint{IPPP/22/23}
\maketitle


\begin{quote}
{\it Give me a lever long enough and a fulcrum on which to place it, and I shall move the world.}

-- Archimedes
\end{quote}

Of the open questions of particle physics and cosmology, the origin of neutrino masses, the baryon asymmetry of the Universe (BAU), and the nature of dark matter (DM) provide perhaps the most well-established evidence for physics beyond the Standard Model (SM).  While the first two involve states and interactions in the SM, it is entirely possible that DM resides in a sector of its own and only indirectly interacts with the known particles.  Nonetheless, most compelling models of neutrino masses
\cite{Minkowski:1977sc,Gell-Mann:1979vob, Mohapatra:1979ia, Yanagida:1979as,Schechter:1980gr} invoke particles -- {\it i.e.} right-handed neutrinos (RHNs) -- that, like DM, have only feeble interactions with the SM.  Remarkably, these right-handed fermions can also provide an interesting resolution of the BAU puzzle through a leptogenesis \cite{Fukugita:1986hr}  mechanism. 

Given the preceding account, it could seem natural to assume that the RHNs and DM are part of a larger ``hidden sector" that is  responsible for the genesis of the ``visible sector" and its large scale structure.  One may then ask if there is a typical energy scale associated with such a hidden sector.  Strictly speaking, there is no robust observational evidence that could provide a clear guide for this question.  Possible mass scales for both RHNs and DM currently span many orders of magnitude.  One is therefore often led to use theoretical motivation in order to arrive at more specific models.

A large class of models focuses on the electroweak scale, where the ``WIMP miracle" (where WIMP stands for weakly interacting massive particle)  motivates cosmologically stable massive particles with weak couplings to the SM.  Furthermore, it is not difficult to imagine that the SM $\vev{H}\approx 246/\sqrt{2}$~GeV \cite{ParticleDataGroup:2020ssz} is itself set by the scale of hidden sector interactions, which could then plausibly be $\sim$ 1-10~TeV.  Connections between such DM candidates and leptogenesis are usually tenuous, as the typical RHN masses are required to be much larger in these scenarios \cite{Davidson:2008bu}.

Based on the above considerations, we will take the point of view that RHNs and DM are from a common hidden sector.  The DM candidate, taken to be a fermion of weak scale mass in what follows, is further assumed to interact with a light scalar that gets displaced far from its origin by the initial thermal population of DM.  This scalar could have additional interactions with the SM, through higher dimensional operators that govern neutrino masses based on a seesaw mechanism.  The framework we will adopt assumes RHNs near the $\sim 10$~TeV mass scale. Interestingly, the light scalar can itself become viable DM, or a component of it, as a result of its displacement, {\it i.e.} a misalignment mechanism.  Since our model is based on lifting parameters through the large  excursion of a scalar, we will refer to it as ``Archimedean Lever Leptogenesis (ALL)."

We will show that the above setup can result in a fleeting enhancement in the interactions of RHNs with the SM, which will eventually fade as the temperature of the Universe and the density of DM fall.  The larger transitory RHN couplings facilitate a viable leptogenesis mechanism around the weak scale, before electroweak symmetry is broken and the processes required to generate the BAU -- {\it i.e.} the electroweak sphalerons \cite{Manton:1983nd,Klinkhamer:1984di} -- are shut off.  At late times, those couplings fall to the levels that are consistent with a neutrino mass seesaw which, barring very degenerate masses for RHNs \cite{Pilaftsis:1997jf,Pilaftsis:2003gt} or SUSY-inspired scenarios with lepton-number violating processes (see e.g.~\cite{Boubekeur:2004ez}), would have been too small to lead to successful leptogenesis.  Our framework thus links the properties of DM with the requirements for successful generation of the BAU.
For recent work in a different context, using a similar mechanism for DM misalignment, see Ref.~\cite{Batell:2021ofv}. Transitory interactions have also been used to modify DM production; see, {\it e.g.}, Refs.~\cite{Cohen:2008nb,Baker:2016xzo,Baker:2018vos,Davoudiasl:2019xeb,Croon:2020ntf}.
We will next introduce a model and the necessary interactions to realize this scenario.

\section{The Hidden Sector}
We will consider a hidden sector that will have suppressed couplings to the SM.  A minimal structure is introduced, since more elaborate assumptions will not affect the main idea in essential ways.   We will assume that the hidden sector includes a real scalar $\Phi$ whose vacuum expectation value (vev) provides mass for the DM fermion $X$.  This fermion carries a chiral $\mathbb Z^{\chi}_2$ parity, with  assignments 
\beq
{\mathbb Z^{\chi}_2} (\Phi)= {\mathbb Z^{\chi}_2} (X_L) = -1 \;\,\text{and}\;\, {\mathbb Z^{\chi}_2} (X_R) = +1, 
\label{Z2chi}
\eeq
with $(L,R)$ denoting (left, right) chirality. To stabilize $X$, we also assume a vector-like parity 
\beq
{\mathbb Z^v_2} (X_R)= {\mathbb Z^v_2} (X_L) = -1 \;\,\text{and}\;\, {\mathbb Z^v_2} (\Phi) = +1. 
\label{Z2v}
\eeq
The RHNs $N_a$, $a=1,2,3$ are assumed to be SM singlets whose masses $M_a\sim 10$~TeV descend from UV dynamics that we shall not specify here. We will also introduce a light real scalar field $\phi$.  The following Yukawa interactions can then be written down
\beq
\mathcal{L} \ni 
\left(y_X + c_X\frac{\phi}{\Lambda_X}\right) \Phi \bar X_L X_R + \sum_{a=1}^3 M_a  \bar N_a^c N_a\,, 
\label{Hidden-Yukawa}
\eeq 
where $c_X$ is a constant taken to be $\ord{1}$. \vtwo{The above dimension-$5$ operator could arise from, for example, a heavy right-handed fermions $\Psi_R$ with the same quantum numbers as $X_R$ and a small coupling to $\phi$ of the type $g_\phi \phi \Psi_R X_R$.
}

The scalar $\Phi$ is assumed to have a simple potential, similar to that of the Higgs field in the SM, realizing $\vev{\Phi} = v_\Phi \neq 0$.  This breaks $\mathbb Z^\chi_2$ and endows $X$ with mass $m_X = y_X v_\Phi$ (at late times when $\phi \to 0$).  We will also take $\phi$ to have an initial mass $m_0$, before electroweak symmetry breaking (EWSB).

Let us now describe how the new scalars $\Phi$ and $\phi$ interact with the SM.  We will start with the scalar potential, including the dim-4 ``portal"  interactions \cite{Patt:2006fw} among the scalars
\beq
V(\phi, \Phi, H) \supset \frac{1}{2} m_0^2 \phi^2 + ( \zeta_\Phi \Phi^2 + \zeta_\phi \phi^2) H^\dagger H\,, 
\label{portal} 
\eeq        
where $\zeta_{\Phi,\phi}$ are constants.\footnote{\vtwo{Note that another portal coupling $\lambda_P \phi^2 \Phi^2$ can be generated at 1-loop through the $X$ coupling. This contribution should at worst be proportional to $ c_X^2 M_{1}^2/\Lambda^2_X$ (where $M_{1}$ is the heaviest state in the effective field theory), which is generically very small in our model.}}  We generally assume that they are both positive.  However, if $\zeta_\Phi <0$ the second term can in principle set the Higgs mass parameter in the SM, with suitable choices of parameters.    This interaction can play a key role in the phenomenology of DM since it allows for $X$ to be in thermal equilibrium with the SM through the coupling of $\Phi$ and $H$.  Also, depending on parameters, the mixing between $\Phi$ and $H$ can provide a channel for direct detection of $X$ through scattering from nucleons mediated by the Higgs boson.   However, in order to keep the analysis simple, we will assume that $\zeta_\Phi$ is sufficiently small so that EWSB largely agrees with the SM expectation.  This implicitly assumes a bare Higgs mass parameter and the required quartic coupling for $H$.  The third term in \eq{portal} will contribute to the mass of $\phi$ after EWSB and can possibly make it much larger than its initial value $m_0$.  In the above setup, we generically have $\zeta_\phi \ll \zeta_\Phi$.   

\section{\boldmath Evolution of $\phi$ with Temperature}
Here, we derive the equation of motion of the scalar $\phi$ in terms of temperature $T$.  Its time evolution is given by
\beq
\frac{d^2\phi}{dt^2} + 3 H \frac{d\phi}{dt}
 + \frac{\partial V(\phi)}{\partial\phi} = 0\,.
\label{eomt}
\eeq
The relevant terms in the scalar potential are $V(\phi) = (m_\phi^2/2) \phi^2 + g_X \phi \bar X X$, where we have defined $g_X \equiv \vev{\Phi}/ \Lambda_X$, and $m_\phi = m_0$ before EWSB.  During radiation domination, we have 
$T=\sqrt{\xi/t}$ and $H(T) = T^2/2 \xi $,
where we have defined
\beq
\xi \equiv \frac{M_P}{2}
\sqrt{\frac{90}{8 \pi^3 g_*(T)}}\, .
\label{xi}
\eeq
with $g_*\sim 100$ the number of the relativistic degrees of freedom and $M_P\approx 1.2 \times 10^{19}$~GeV the Planck mass.

We assume the dark sector Higgs mechanism takes place at $T_X\gsim 100$~GeV, giving the DM state $X$ a mass $m_X \sim 100$~GeV for $\vev{\Phi}\gsim 100$~GeV.
The portal interaction between $\Phi$ and the Higgs can thermalize $\Phi$ and hence $X$ with the SM, setting up the initial conditions for a thermal relic DM scenario. In the thermal bath, $\bar XX$ acts like the following Lorentz invariant expression,
\beq
\bar X X\to n_X \vev{\sqrt{1 - v_X^2}}\,,
\label{eq:XbarX}
\eeq
where $v_X$ is the speed of $X$ and $n_X$ is its number density.  

One can then straightforwardly show that the evolution of $\phi$ with $T$ is governed by 
\beq
\frac{T^6}{4 \xi^2}\frac{d^2\phi}{dT^2} + 
m_\phi^2 \phi + g_X n_X \vev{\sqrt{1-v_X^2}}= 0\,.
\label{Tevol}
\eeq
To proceed, note that effective mass of $X$ is $\mbX \equiv m_X + g_X \phi$,  related to its energy via $E_X = \mbX/\sqrt{1-v_X^2}$ and hence
\beq
\vev{\sqrt{1-v_X^2}} = \mbX \vev{\frac{1}{E_X}}.
\label{avinvboost}
\eeq

Since $X$ is a Dirac fermion, its thermal distribution is given by $f(p) = (e^{E/T} + 1)^{-1}$, assuming zero chemical potential (which is a good approximation in our scenario), $E$ is energy and $p$ denotes momentum.  We have
\beq
\begin{split}
\vev{\frac{1}{E_X}} &= \frac{g}{(2 \pi)^3 n_X}\int E_X^{-1} f(p_X) d^3p_X\,
\\ &=  
\frac{g}{ \vtwo{2 \pi^2} n_X}
\int f(p_X)\sqrt{E_X^2 - \mbX^2} dE_X,
\end{split}
\label{avinvE2}
\eeq
where 
\beq
\begin{split}
n_X &= \frac{g}{(2 \pi)^3}\int f(p_X) d^3p_X \\ 
&= \frac{ g}{\vtwo{2\pi^2}}\int f(p_X)\sqrt{E_X^2 - \mbX^2} \, E_X dE_X,
\label{nX}
\end{split}
\eeq
and we assume $g=4$ for $X$ and $\bar X$.  
We now note that the expression for the pressure $P_X$ is
\beq
P_X = \frac{ g}{\vtwo{6
\pi^2}}\int f(p_X)\left(E_X^2 - \mbX^2\right)^{3/2} dE_X\,,
\label{PX}
\eeq
which together with \eq{avinvE2} implies 
\beq
\vev{\frac{1}{E_X}} = \left(\frac{-1}{\mbX n_X}\right)
\frac{\partial P_X}{\partial \mbX}.
\label{avinvE3}
\eeq
A similar expression was found in Ref.~\cite{Domenech:2021uyx} from the fundamental thermodynamic relations at constant particle number and temperature.

One can find an expansion for $P_X$ in $\mbX/T$ (see, for example Ref.~\cite{Domenech:2021uyx}) for a relativistic thermal population 
\beq
\frac{P_X }{T^4} \approx   \frac{1}{3}\left(\frac{7}{8}\right)\left(\frac{\pi^2}{30}\right) g 
- \frac{g}{48} \frac{\mbX^2}{T^2}
+ \ord{\frac{\mbX^3}{T^3}}.
\label{PXexpansion}
\eeq
As the evolution of $ \phi$ reduces $ \mbX $, this is a good approximation if $m_X \ll T $ when $X$ first gets a mass. 
Note that the leading term gives $P_X = \rho_X/3$, with $\rho_X$ the energy density of radiation made of $(X,\bar X)$.  
Putting the above together, we get
\beq
\vev{\frac{1}{E_X}} = 
\frac{T^2}{n_X}\left[\frac{g}{24} + \ord{\frac{\bar m_X}{T}}\right].
\label{avinvE4}
\eeq
This result implies that in the limit $ \bar{m}_X \to 0$, the fermion $X$ behaves like pure radiation and its effect on the $\phi$ equation of motion vanishes.
From \eqref{Tevol}, \eqref{avinvboost}, and \eqref{avinvE4}, the evolution of $\phi$ with temperature is given by
\beq
\frac{T^6}{4 \xi^2}\frac{d^2\phi}{dT^2} + 
\left(m_\phi^2 + \frac{g_X^2}{6}\,T^2 \right) \phi + \frac{g_X}{6}\, m_X T^2 = 0.
\label{Tevolrel}
\eeq

The above equation leads to different behaviors for $\phi$ depending on the relative importance of various terms.  When the Hubble scale is larger than the effective scalar mass -- that is, both the thermal contribution and initial $\phi$ mass -- the evolution is driven by the first and the last terms and 
\beq
\frac{d^2\phi}{dT^2}\sim - g_X m_X H(T)^{-2} \; \Rightarrow \; \phi \sim - \frac{g_X m_X \xi}{H(T)}
\propto - t\,,
\label{phi-H-dom}
\eeq
and roughly grows with time, assuming it starts out with vanishing initial velocity and field value.

\begin{figure}
    \centering
    \includegraphics[width=.49\textwidth]{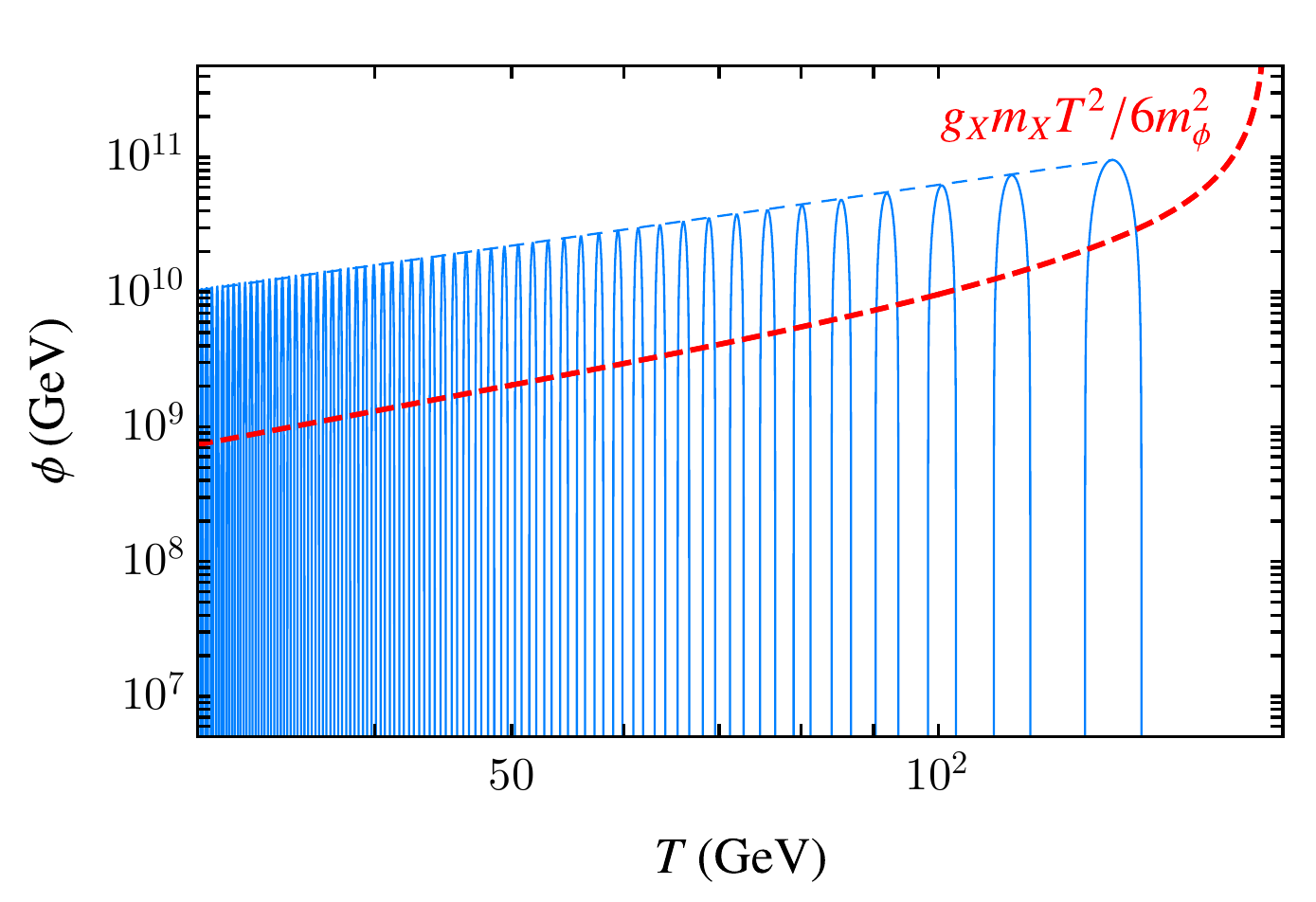}
    \caption{
    Behavior of $\phi$ as a function of temperature (in blue) and its envelope (in dashed blue). In this plot, the X mass switches on at $T_X = 150$ GeV with a value of $m_X = 50$ GeV, and during EWSB $\phi$ develops a mass $ m_\phi = 10^{-3}$ eV ($m_\phi = 10^{-5}$ eV before EWSB). We have assumed $g_X = 10^{-19}$ and $g_* = 110$. This leads to $\Omega_\phi \sim 0.10 \, \Omega_{\rm DM} $.
    }
    \label{fig:phiplot}
\end{figure}

Once the Hubble scale is not dominant compared to the effective $\phi$ mass, the solution 
\vtwo{will oscillate around}
\beq
\phi \sim 
-\frac{g_X T^2 m_X}{6 m_0^2 + g_X^2 T^2}\,.
\label{phi-mass-dom}
\eeq
When the $g_X^2 T^2$ term dominates, \vtwo{\eqref{phi-mass-dom}} tends to $\phi \to -m_X/g_X$, 
until $T$ has become sufficiently small, or else there is a jump in $m_0$.  However, if $m_0^2$ is dominant then this ``attractor solution" is not reached and $\phi$ assumes a value given by  
\beq
\phi \sim 
-\frac{ g_X T^2 m_X}{ 6 m_0^2  }.
\label{phi-m0-dom}
\eeq
We demonstrate this behavior in Fig.~\ref{fig:phiplot}.  Here we have modeled EWSB in the high temperature expansion of the Higgs potential (see, {\it e.g.}, Ref.~\cite{Quiros:1999jp}), valid until $T \sim 50$ GeV, like the approximation \eqref{PXexpansion} for $m_X = 50$~GeV. We find that the results can vary slightly depending on how the Higgs vev switches on; the dynamics is dominated by the higher temperatures. In particular, we have checked that $\Omega_\phi$ scales as $T^3$ before the breakdown of the high temperature expansion. We have modeled the dark sector Higgs mechanism in a similar fashion, but at higher temperatures: at $T \sim 150$ GeV, the $X$ mass has reached its $T=0$ value, such that the precise dynamics of this phase transition are unimportant. 

We conclude this section by briefly remarking on the non-relativistic limit of the scalar equation of motion, which may become important for alternative model parameters (in the regime $T\lsim m_X/3$). In this limit, 
an expansion in $g_X \phi/T\ll 1$ and $g_X \phi/m_X\ll 1$, which is valid for the parameters we will consider, yields
\beq
\frac{T^6}{4 \xi^2}\frac{d^2\phi}{dT^2} + 
\left[m_\phi^2 - \frac{g_X^2 n_{0X}}{T}\left(1-\frac{3 T}{2 m_X}\right)
\right] \phi + g_X n_{0X} = 0,
\label{Tevolnonrel}
\eeq
where 
\beq
n_{0X} = g\, (2\pi)^{-3/2} (m_X T)^{3/2} e^{-m_X/T}.
\label{n0X}
\eeq
Note that $ n_{0X}$ depends on $m_X$, whereas $n_X$ in \eq{nX} depends on $ \bar{m}_X$.

\section{DM Support for Leptogenesis} 
\begin{figure*}[t!]
    \centering
    \includegraphics[width=0.8\textwidth]{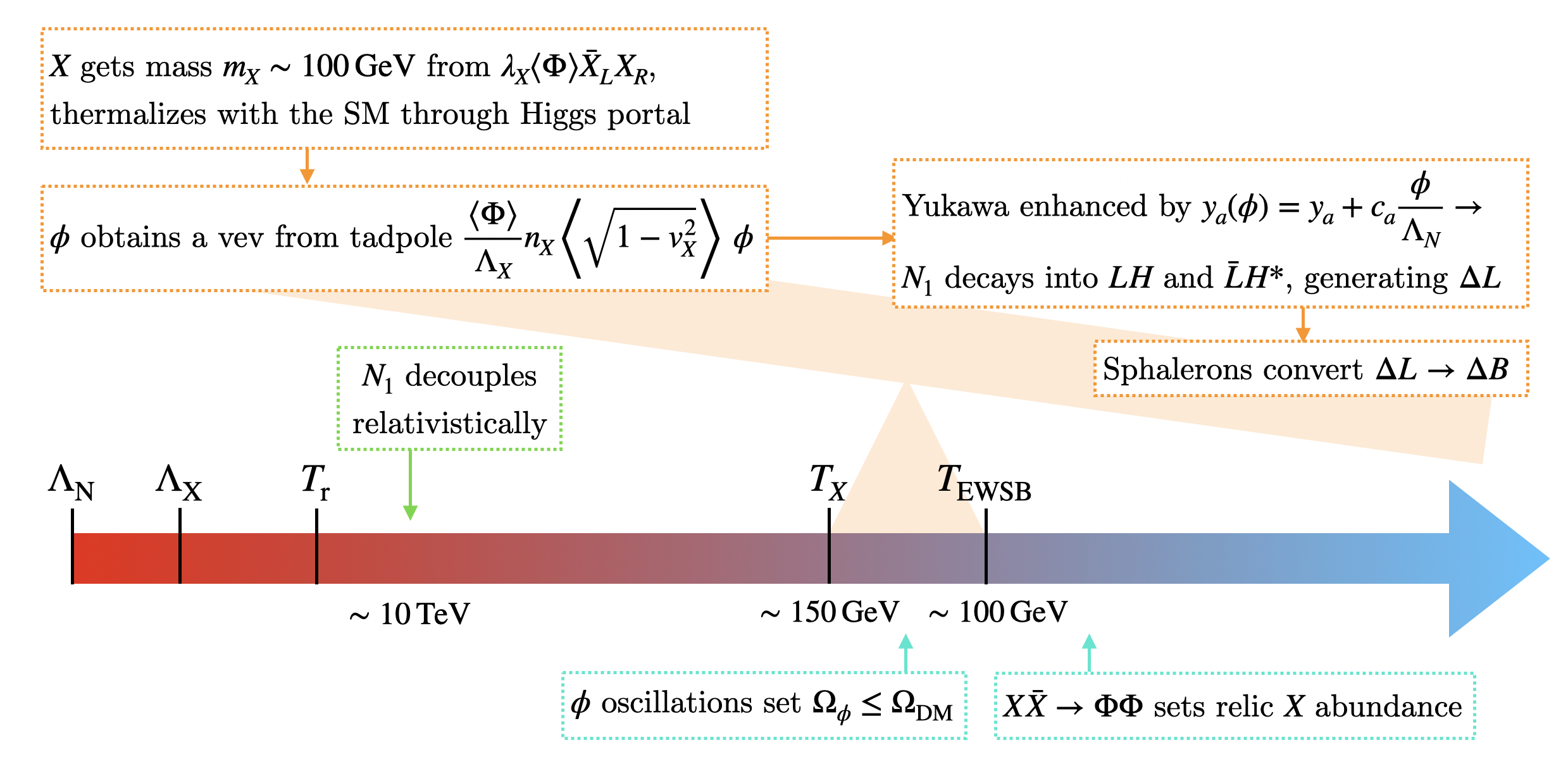}
    \caption{Schematic illustration of ALL.
    }
    \label{fig:schematic}
\end{figure*}

We will now demonstrate that the scenario described above lends itself to leptogenesis at the weak scale.\footnote{For a schematic illustration of the mechanism, see Fig.~\ref{fig:schematic}.}
Let us assume that there are 3 right-handed neutrinos $N_a$, $a=1,2,3$, of masses $M_a$.  There could be a mild hierarchy of masses, but we will generally assume $M_a\sim 10$~TeV $\forall a$.  To get the experimentally implied SM (left-handed) neutrino mass $m_\nu \lsim 0.1$~eV, we consider a seesaw, provided by the Dirac mass terms
\beq
\sum_{a,i=1}^3 \left(y_{ai} + c_{ai} \frac{\phi}{\Lambda_N}\right) \bar N_a H \epsilon L^i + {\small \text{H.C.}}\,,
\label{Dirac}
\eeq 
where $\Lambda_N$ is a high UV scale and $y_{ai},c_{ai} \in \mathbb C$. 
\vtwo{This operator may arise in a similar effective field theory (EFT) as \eqref{Hidden-Yukawa}. We note in passing that while some of the values of the couplings in our EFT are quite small, they are stable against quantum corrections in our model.}
To keep the $N_1$ population -- which is assumed to be generated at $T\gg 100$~GeV -- from decaying away, we require that $y_{1i}=0$, or else sufficiently tiny.   

Note that the largest typical\footnote{\vtwo{E.g. without assuming some particular texture for the Yukawa matrices, which would require a cancellation to reproduce $m_\nu$.
}} value of $y_{ai}$ is given by 
\beq
y_{ai}  \lsim 10^{-5} \left(\frac{m_\nu}{\text{0.1 eV}}\right)^{1/2} \left(\frac{M_a}{\text{10 TeV}}\right)^{1/2}.
\label{yai}
\eeq  
We will see that the above values of $y_{ai}$ are too small to obtain a sufficient amount of baryon asymmetry from decays of $N_a$, 
for $M_1 \sim 10$~TeV.  However, the $\phi$ dynamics described in the previous section gives rise to Yukawa couplings at $T\sim 100$~GeV and a viable leptogenesis mechanism even if $y_{ai}$ are zero when $\vev{H}=0$.  Since we will assume that the enhanced transitory couplings $\propto c_{ai}\phi$ will be much larger than those at $T=0$, we will not specify the form of the $y_{ai}$ matrix that can lead to realistic phenomenology at late times and the observed properties of neutrinos.  

We assume that $N_1$ established thermal equilibrium with the SM, through scattering mediated by a heavy scalar $S$ (with mass $\gg 10$~TeV) down to $T\gsim 10$~TeV.  One can easily arrange for $S$ to have small couplings to the SM such that $N_1$ gets  decoupled while still relativistic, maintaining a number density $\sim T^3$ (akin to SM neutrinos that decouple for $T\lsim$~MeV).

The scalar $\phi$ will not be in thermal equilibrium during leptogenesis, since we will assume that it has sufficiently small interactions.  Production rates that scale like $\sim T$ will {\it recouple} at low temperatures, so we need to ensure they are ineffective at the lowest temperature of interest, which is near $T_*\sim 100$~GeV of EWSB.  If this is ensured they will remain decoupled at higher $T$.  This roughly requires $\zeta_\phi^2 T_*$ and $g_X^2 T_*$ to be small compared to Hubble rate  $H(T_*)\sim\sqrt{g_*}\, T_*^2/M_P \sim 10^{-14}$~GeV.  We hence require $g_X, \zeta_\phi \lsim 10^{-8}$.  

Let us take $g_X \sim 10^{-20}$, which for $T\sim m_X\sim 100$~GeV and $m_0< H(T)$ yields $\phi\sim 10^{11}$~GeV.  This agrees with the plotted behavior of $\phi$ in Fig.~\ref{fig:phiplot}.  Note that the plot assumes that the final mass of $\phi$ is $m_\phi=10^{-3}$~eV, 
which is larger than the above $m_0\lsim 10^{-4}$~eV.  We have assumed  that this is because the Higgs vev $v_h\approx 246$~GeV makes a contribution to $m_\phi^2$ of order $\zeta_\phi v_h^2$, implying that $\zeta_\phi \sim m_\phi^2/v_h^2\sim 10^{-28}$. Hence $g_X$ and $\zeta_\phi$ are consistent with our assumptions above on the upper bound on these couplings.

For rates that grow faster than $T^2$, we need to ensure they are decoupled at the {\it highest} relevant $T$, since they will then remain decoupled as $T$ drops.  We would then consider the Hubble rate  $H(T_{UV}) \sim 10^{-8}$~GeV at $T_{UV}\sim 100$~TeV, possibly the UV regime where the $N_a$ population originates form.  Hence, the  production rate needs to be small compared to $H(T_{UV})$.  So, we would take $T\sim T_{UV}$ for dimension-5 operators in \eq{Dirac}, corresponding to the rate $\sim T_{UV}^3/\Lambda_N^2$.  For the reference parameters above we will find that $\Lambda_N\sim 10^{15}$~GeV is a typical value for our scenario, which will keep $\phi$ out of thermal contact with the SM.

We will assume that the lepton asymmetry $\Delta L$ is generated through the decays $N_1 \to L H$ and $N_1 \to \bar L H^*$, with partial widths $\Gamma$ and $\bar \Gamma$.  Let us define the asymmetry parameter 
\beq
\eps \equiv \frac{\Gamma - \bar \Gamma}{\Gamma + \bar \Gamma} 
\label{eps-def}
\eeq  
and the $\phi$-dependent Yukawa couplings 
\beq
y_a(\phi) = y_a + c_a \frac{\phi}{\Lambda_N}.
\label{ya}
\eeq
We have suppressed the lepton generation index $i$ in the above and what follows, taking them to be of similar size for each RHN.  
The numerator of \eq{eps-def} $\sim |y_1(\phi)|^2|y_{2,3}(\phi)|^2 \sin \theta$, where $\theta\neq 0$ is the physical phase associated with CP violation in the Yukawa couplings.  The denominator of $\eps$ is dominated by the tree-level decay processes for $N_1$ which is $\sim |y_1(\phi)|^2$.  For simplicity, let us take $M_2=M_3=M_N$ and $y_2(\phi) = y_3(\phi) = y_N(\phi)$.  Assuming that $|y_1(\phi)|\ll |y_N(\phi)|$, as we will do below, then we roughly get \cite{Davidson:2008bu} 
\beq
\eps \sim \frac{3}{8\pi}\frac{M_1}{M_N}|y_N(\phi)|^2 \sin\theta,   
\label{eps-val}
\eeq
for a mild hierarchy $ M_1 < M_N$.

Leptogenesis begins once $m_X\neq 0$ at $T\sim T_X$, the scalar $\phi$ gets a tadpole vev and leads to enhanced couplings of $N_a$ to Higgs and leptons.  Since $N_{2,3}$ are assumed to have the required couplings for a viable seesaw at $T=0$ from the start, they will have been efficiently depleted from the plasma.  As stated before, we will assume that $N_1$ has negligible coupling to $H\,L$ and hence its population does not decay, once produced at $T\gg T_X$.  The $N_1$ population however must quickly decay once $\phi$ has enhanced its Yukawa couplings in order to generate a lepton asymmetry $\Delta L$.  The electroweak sphalerons turn $\Delta L$ into a baryon asymmetry $\Delta B$ before EWSB. 

In order to achieve leptogenesis, we need the $N_1$ population to decay away before EW symmetry is broken at $T_*\sim 100$~GeV and the sphaleron processes are shut off.  We then roughly require that the width of $N_1$ exceed the Hubble rate at $T_*$,
\begin{equation}
    \Gamma(N_1)\sim \dfrac{y_1^2(\phi)}{16 \pi} M_1 > H(T_*) \sim 10^{-14} \, \rm{GeV},
\end{equation}
which implies
\beq
y_1(\phi)\gsim 10^{-8}, \quad\text{($N_1$ decay before EWSB)}
\label{N1decay}
\eeq
which can easily accommodate the requirement on asymmetry ``washout" via $N_a$ exchange, as explained below.

One may ask whether the requisite $y_1(\phi) \gsim 10^{-8}$ obtained above may imply a fast three-body decay of $N_1 \to \phi H L$ before $T\sim T_X$, removing the $N_1$ population before the enhanced couplings  necessary for leptogenesis are achieved.  Based on the preceding analysis, let us take a ``safe" value $y_1(\phi) \approx c_{ai} \,\phi/\Lambda_N \sim 10^{-7}$ from \eq{ya}.  For the typical value $\phi \sim 10^{11}$~GeV adopted before in our discussion, we then have $\Lambda_N/c_{ai}\sim 10^{18}$~GeV.  One can estimate the three-body decay mediated by the dimension-5 operator in \eq{Dirac} to give a rate $\ll |c_{ai}|^2M_1^3/\Lambda_N^2 \sim 10^{-24}$~GeV which is much smaller than the Hubble scale at $T > T_X$.

To determine parameters that avoid washout of the asymmetry generated by $N_1$ decay, let us consider dim-5 operators 
\beq
O_a=\frac{|y_a(\phi)|^2 (H\epsilon L)^2}{M_a}.
\label{dim5}
\eeq
obtained by integrating out $N_a$, as they are heavy compared to $T_X$ and their production is suppressed by $e^{-M_a/T_X}$ with $M_a/T_X\sim 100$.  The rate $\Gamma_W$ of the processes mediated by $O_a$ should be smaller than the Hubble rate $H(T_X)\sim 10^{-14}$~GeV.  For the washout, $\Gamma_W \sim |y_a(\phi)|^4 (T_X^3/M_a^2)$, and we would need $\Gamma_W <  H(T_X)$.  Hence, for $N_a$ we get
\beq
y_a(\phi)\lsim 10^{-3},\quad\text{(Ineffective Washout)} 
\label{Washout-bound}
\eeq
for similar $M_a$ at $T\sim T_X$.  This upper bound allows a broad range of values for $y_a$.

Let us now estimate the minimum value for $y_N(\phi)$ to generate $n_B/s \sim 10^{-10}$ \cite{ParticleDataGroup:2020ssz}, where $n_B$ is the baryon number density and $s\sim g_* T_X^3$ is the entropy density.  If the initial population of the $N_1$ is relativistic, its number density is given by $n_1\sim T_X^3$.  Then, one finds 
\beq
\frac{n_B}{s} \sim \frac{n_1 \, \eps}{s}\sim \frac{3}{8\pi g_*}\frac{M_1}{M_N}|y_N(\phi)|^2 \sin\theta.
\label{nB}
\eeq
In the above, we have ignored $\ord{1}$ coefficients that relate baryon and lepton asymmetries via sphaleron processes.  For $\sin \theta\lsim 1$ and $M_1/M_N\sim 1$, we then find
\beq
y_N(\phi_{T_X})\gsim 3\times 10^{-4},\quad\text{(Sufficient BAU)} 
\label{BAUbound}
\eeq
which is consistent with the washout upper-bound on this coupling from the preceding discussion.
Note that the $y_N$ is smaller than in standard leptogenesis scenarios, because $N_1$ decouples relativistically and thus there is no Boltzmann suppression. Typical values of $\phi$ today easily accommodate \eqref{yai}.

We can generate a non-zero $\phi$-independent value for $y_1$ after EWSB by introducing the higher dimension operators $\propto H^\dagger H$, while avoiding fast multi-body decays of $N_1$, though this is not going to change our basic scenario in any important ways.  Hence, we take the simple implementation above that implies one of the SM neutrinos is much lighter than the other two, since there is effectively only a $2\times 2$ Dirac mass matrix in \eq{Dirac}.

\section{Thermal Relic Dark Matter}
\begin{figure}[t]
    \centering
    \includegraphics[width=.49\textwidth]{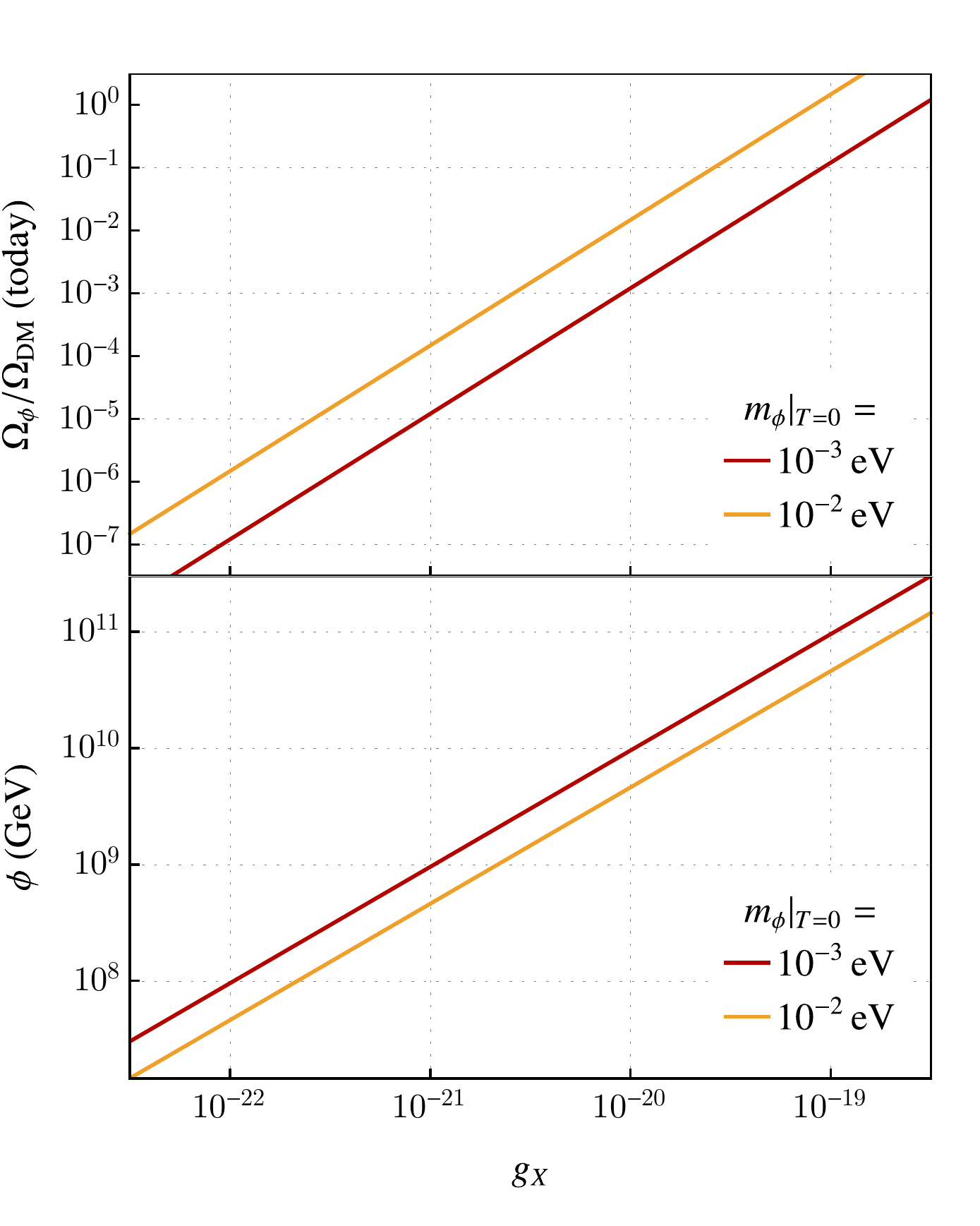}
    \caption{
    Sub-fraction of dark matter constituted by $\phi$, and initial displacement of $ \phi$.
    Here we have assumed that $m_X|_{T=0} = 50$~GeV, $g_* = 110$, and that $m_\phi = 10^{-5}$ eV before EWSB.    }
    \label{fig:combined}
\end{figure}

We will now show that both $\phi$ and $X$ can play the role of dark matter in this model. Let us first show an example in which the final $\phi$ abundance is subdominant, and $X$ plays the role of dark matter. This scenario is realized for the parameter values in Fig.~\ref{fig:phiplot}: we find the final $\phi$ abundance to be $\Omega_\phi \sim 10^{-1} \Omega_\text{DM}$ (with $\Omega_\text{DM}=0.27$ \cite{ParticleDataGroup:2020ssz}). See Fig.~\ref{fig:combined} for the final $\phi$ abundance for different values of $g_X$ and $m_\phi$ after EWSB.

In order to simplify the treatment we consider a scenario in which $X$ maintains its thermal abundance until {\it after} EWSB at $T_* \approx 100$~GeV.  Hence, we will assume that $m_X\lsim T_*$.  Furthermore, we will require $m_X > m_\Phi$ such that $X \bar X \to \Phi \Phi$ can set the relic abundance.  In this case, we will need to assume that $\Phi$ would decay into SM states at some point, but this will not impose any severe  restrictions on our model.  In principle, if $\Phi$ couples to the SM with appropriate strength it could potentially lead to signals in high energy experiments.  

We will take the $\Phi$ potential to be of the form
\beq
V_\Phi = -\frac{1}{2} m_\Phi^2 \Phi^2+ \frac{\lambda_\Phi}{4!} \Phi^4\,,
\label{VPhi}
\eeq
which would give
\beq
v_\Phi = \sqrt{\frac{6}{\lambda_\Phi}} m_\Phi,
\label{vPhi}
\eeq
where $m_\Phi$ is the mass of $\Phi$, after the phase transition that breaks the assumed $\mathbb Z^\chi_2$, via $v_\Phi\neq 0$.  Since we have  $m_X = y_X v_\Phi$, our condition $m_X>m_\Phi$ on masses yields \beq
\lambda_\Phi < 6 y_X^2.
\label{lamPhi-upper}
\eeq
For a choice of $m_X$, one can find a value for $y_X$ that results in the right relic abundance through annihilation.  Then, one has to assume that $\lambda_\Phi$ satisfies \eq{lamPhi-upper}, for consistency.

Let us denote the temperature at which $\Phi$ gets a vev and $X$ becomes massive by $T_X$ and assume, for simplicity, that $\lambda_\Phi \ll 4 y_X^2$.  One can show (see, for example, Ref.~\cite{Davoudiasl:2015vba}) 
\beq
T_X \approx \frac{\sqrt{6}}{y_X} m_\Phi.
\label{TPhi}
\eeq
Since we want $m_X < T_*$ and $T_X > T_*$, so that leptogenesis occurs when sphalerons are still active, we have 
\beq
\lambda_\Phi = r^2 y_X^4\quad ; \quad 
r > 1\,,
\label{TPhi-lower}
\eeq
where $r\equiv T_X/m_X$.  The above, together with \eq{lamPhi-upper}, yields 
$y_X \lsim \sqrt{6}/r$.  Note that the limit assumed in deriving \eq{TPhi} implies $r^2 y_X^2 \ll 4$.  Hence, we require 
\beq
r \, y_X \lsim 1\,,
\label{ryX}
\eeq
as a consistency condition on our parameters.  In summary, the choice of $m_X$ fixes $y_X$, subject to \eq{ryX}, together with $T_X > T_*>m_X$, so that the above DM scenario can be realized.

The cross section for the annihilation of $X$ through a scalar mediator $\Phi$ is given by (see, for example, Ref.~\cite{Kaplinghat:2013yxa})
\beq
\sigma_X v = \frac{3\, v^2 y_X^4}{128 \pi\, m_X^2}\,,
\label{sigmaXv}
\eeq
where $v$ is the relative velocity of $X$ and $\bar X$.  Using $\vev{v^2} = 6 T/m_X \approx 0.32$, for $p$-wave suppression relevant to $m_X\lsim 100$~GeV in our work, we find for the $X$ energy density
\beq
\frac{\Omega_X}{0.27} \approx  \frac{4.4\times  10^{-26}~\text{cm}^3{s}^{-1}}{\vev{\sigma_X v}} \approx 
\left(\frac{0.25}{y_X}\right)^4\left(\frac{m_X}{50~\text{GeV}}\right)^2.
\label{OmegaX}
\eeq
Hence, we find that for $m_X\sim 50$~GeV and $r\sim 3$, corresponding to $T_\Phi\sim 150$~GeV, we can realize the DM scenario sketched above.

We may also consider the scenario in which $\phi$ plays the role of DM. This scenario can be realized for a small modulation of our parameters and possibly lead to a multi-state DM sector, if we maintain $X$ as one of its major components.  Alternatively, one may also arrange for $\phi$ to be the dominant DM; 
sample parameters for $\phi$ DM can be inferred from Fig.~\ref{fig:combined}. The field $X$ can then be sufficiently depleted, through the Higgs portal,  for somewhat larger annihilation cross section than required for DM (otherwise, one may  arrange for $X$ to decay away after electroweak symmetry breaking).   

\section{Phenomenology} 
Let us now discuss some of the potential experimental consequences of the scenario described above.  We note that the details of the phenomenology depend of the benchmark parameters.  However, there are a number of general possibilities that can arise in our framework.  First of all, since we have assumed that $\Phi$ and the Higgs can potentially mix, one could provide a path for dark matter $X$ to couple to the SM.  This will allow its freeze-out relic density to be set, yet the magnitude of coupling could be small, as we have assumed a ``light mediator" mechanism through annihilation into $\Phi\Phi$. 
We also generally assumed that the Higgs mixing with $\Phi$ is not large to avoid changing the SM EWSB phase transition.  Nonetheless, one could in principle consider versions of our model where this mixing is significant.

To examine the  phenomenological implications of $H$-$\Phi$ mixing, let us first
estimate the minimum level of interaction between the Higgs and $\Phi$ necessary to thermalize the latter.  For values of $y_X\sim 0.25$ near the benchmark adopted above, the $\Phi$ Yukawa coupling to $X$ will then bring $X$ into equilibrium with the SM prior to its freeze-out.  Before EWSB, the the portal coupling of $\zeta_\Phi \Phi^2 |H|^2$ in \eq{portal} can lead to thermalization of $\Phi$, as long as $\zeta_\Phi^2 \gsim g_*^{1/2} T_*/M_P$, which implies
\beq
\zeta_\Phi\gsim 10^{-8},\quad
\text{(Requirement for ALL)}.
\label{zetamin}
\eeq
This 
easily avoids any conflict with  current constraints, as will be discussed below.

The $\Phi$-$H$ mixing in our model is governed by the angle 
\beq
\Theta \approx \frac{2 \zeta_\Phi v_h v_\Phi}{m_H^2}\,,
\label{delta}
\eeq
for $m_\Phi^2\ll m_H^2$, where $m_H\approx 125$~GeV is the observed Higgs mass \cite{ParticleDataGroup:2020ssz} and $v_\Phi\equiv \vev{\Phi}$.  
Adopting the benchmark values of parameters employed in the preceding discussion, corresponding to $m_\Phi\sim 15$~GeV, we have $v_\Phi\sim 200$~GeV, which we will use in what follows.  Hence, we have $\Theta \sim 6\times \zeta_\Phi$. We first consider the case that $X$ makes up all of dark matter. Using the results of Ref.~\cite{Kaplinghat:2013yxa}, the spin-independent $X$-nucleon scattering cross section, mediated by $\Phi$, is estimated to be 
\beq
\sigma_{Xn}\sim 2\times 10^{-40}~{\rm cm}^2\, \Theta^2 \left(\frac{y_X}{0.25}\right)^2 \left(\frac{15~{\rm GeV}}{m_\Phi}\right)^4.
\label{sigXN}
\eeq
For $m_X\sim 50$~GeV, as chosen in the above discussion of $X$ thermal relic abundance, the current bound from the Xenon1T experiment is $\sigma_{Xn}\lsim 5\times 10^{-47}$~cm$^2$ at 90\% CL \cite{XENON:2018voc}, which implies $\Theta\lsim 5 \times 10^{-4}$ and hence 
\beq
\zeta_\Phi\lsim 10^{-4}\,,\quad \text{(Direct Detection; $m_X\sim 50$~GeV)}  
\label{zetaDMDD}
\eeq
assuming that DM is all composed of $X$.  We see that this upper bound is about 4 orders of magnitude above the minimum $\zeta_\Phi$ required for thermalization of $X$, derived before.

Next, we consider bounds that apply when $X$ does not necessarily make up the dominant component of dark matter. The $\Phi$-$H$ portal also allows $H\to \Phi \Phi$.  The width for this decay is given by \cite{Kaplinghat:2013yxa} 
\beq
\Gamma (H\to \Phi \Phi)\approx \frac{\zeta_\Phi^2 v_h^2}{8\pi\, m_H}\,,
\label{H2Phi}
\eeq
for $m_\Phi\ll m_H$. 
Assuming the width of the Higgs is approximately the same as in the SM, $\sim 4$~MeV \cite{LHCHiggsCrossSectionWorkingGroup:2013rie}, which is a consistent assumption here, we find the corresponding branching ratio
\beq
\text{Br}(H\to \Phi \Phi)\lsim 5 \times 10^3 \zeta_\Phi^2\ .
\label{BrH2Phi}
\eeq
Since $m_\Phi < m_X$ in our scenario, the main decay channels of $\Phi$ are those accessible through mixing with the Higgs.  
For $ 12\, \text{GeV} \lesssim  m_\Phi \lesssim m_H/2$,
this means that dominant decay channel of $\Phi$ is into $b$ quark pairs.  We expect $\sim 80\%$ for the branching ratio of $\Phi \to b\bar b$, with the rest mostly shared among gluon, $\tau$, and charm quark pairs, as may be approximately deduced from the Higgs branching fractions in the SM \cite{LHCHiggsCrossSectionWorkingGroup:2013rie}. 

For experimental bounds, we note that the decay width of $\Phi\to b\bar b$ is of order $\Gamma_\Phi\sim \Theta^2 (m_b^2/v_h^2) m_\Phi$, where $m_b\sim 4$~GeV is the $b$ quark mass \cite{ParticleDataGroup:2020ssz}.  
We will use the ATLAS search results for Higgs decay into a pair of scalars that each promptly decay into $b \bar b$ \cite{ATLAS:2018pvw}, which is the same process we have in our scenario. This search focuses on Higgs production in association with a $W$ or $Z$ boson, which have SM NNLO cross sections $1.37$~pb and $0.88$~pb, respectively \cite{LHCHiggsCrossSectionWorkingGroup:2013rie}.  The ATLAS upper bound on the product of  combined production cross section times $\text{Br}(H\to \Phi \Phi\to 4 b)$, at the 13~TeV LHC  with 36.1 fb$^{-1}$, is $\sim 1.25$~pb, assuming a $\sim 40$~GeV scalar (at 95\% CL).  
This implies
\beq
\zeta_\Phi < 1 \times 10^{-2} \,,\quad ( H\to \Phi \Phi),
\eeq
which is clearly only relevant if $X$ is not the DM.  
At the above upper limit 
we have $\Gamma_\Phi\sim 10^{-5}$~GeV.  This value of $\Gamma_\Phi$ corresponds to a $\Phi$ decay length  $\ll\mu$m, a posteriori motivating our assumption of promptness \cite{ATLAS:2018pvw}.

Assuming $\sim 100$ times more data by the end of the LHC high luminosity operations, if $X$ is a significant component of DM, we still do not expect sensitivity to our range of parameters, which is much more stringently constrained by direct detection bounds.  The preceding analysis, incidentally, implies that even for the minimum $\zeta_\Phi \sim 10^{-8}$ we will have $\Gamma_\Phi\sim 10^{-17}$~GeV, which corresponds to the Hubble scale at $T\sim$~GeV, allowing $\Phi$ to decay well before the BBN. 

The LHC could also potentially probe our scenario through invisible Higgs decays $H\to X\bar X$, assuming $m_X<m_H/2$,  
with a rate \cite{Kaplinghat:2013yxa} \beq
\Gamma(H\to X\bar X) = \frac{y_X^2}{8\,\pi} \Theta^2 m_H
\left(1- \frac{4 m_X^2}{m_H^2}\right)^{3/2}.
\label{GammaHXbarX}
\eeq
Note that if $X$ is DM, \eq{zetaDMDD} implies $\Theta \lsim 5\times 10^{-4}$ and hence we expect the above decay to have a branching fraction $\lsim 10^{-5}$, which is well below the current LHC constraints $\lsim 0.2$ \cite{CMS:2022qva,ATLAS:2022yvh} and foreseeable ones.  If $X$ is not DM, the constraint is given by
\begin{equation}
    \zeta_\Phi \lsim \frac{2 \times 10^{-3}}{y_X}, \quad (H \to \bar{X}{X}),
\end{equation}
assuming $ v_\Phi = 200$~GeV as before.

Another possible signal of our framework is the emergence of a long-range force mediated by the light scalar $\phi$.  Here, one route for linking $\phi$ to the SM is through quantum processes involving an $X$ loop that connects $\phi$ and $\Phi$, and hence to the Higgs through $H$-$\Phi$ mixing.  However, as mentioned before this mixing could be small and the coupling of $\phi$ to $X$ is also generally tiny in our model.  So, this may not be a typical path for $\phi$ to interact measurably with the SM baryon and charged leptons.  We, therefore, focus on the  couplings of $\phi$ given in \eq{Dirac}, in the following.

The typical size of $H$ Yukawa couplings in \eq{Dirac}, using our benchmark model parameters, is given by $y_N \sim 10^{-5}$, for $N_{2,3}$ states.  However, the coupling of $\phi$ to $L N_a$ depends on its initial amplitude, since we would like to have $c_{ai} \phi/\Lambda_N\sim 10^{-4}$.  As an example, let us take the mass of $\phi$ after EWSB to be $m_\phi\sim 10^{-3}$~eV and its initial value $\phi\sim 10^{11}$~GeV, as adopted before in our discussion.  Assuming $|c_{ai}|\sim 1$, we then have $\kappa \sim |c_{ai}| \vev{H}/\Lambda_N\sim 10^{-13}$ which sets the $T=0$ coupling of $\phi$ to $LN_a$ with $\Lambda_N\sim 10^{15}$~GeV.  We then estimate that the 1-loop coupling of $\phi$ to $t\bar t$ is given by
\beq
g_{\phi t}\sim \frac{\kappa\, y_t \, y_N M_N^2}{16 \pi^2 m_H^2}\sim 10^{-16}\,,
\label{gphit}
\eeq
where $y_t\approx 1$ is the SM top Yukawa coupling. 

The coupling $g_{\phi t}$ can be translated into a coupling to nucleons $g_{\phi n}$, where $g_{\phi n} \sim 10^{-3} g_{\phi t}$ \cite{Knapen:2017xzo}.  For $m_\phi \sim 10^{-3}$~eV, this value of $g_{\phi n}$ is just inside the region excluded by tests of the inverse square law \cite{Adelberger:2009zz,Heeck:2014zfa}.  Hence, we conclude that current tests of new long range forces and their improvements could probe our setup for parameters near what has been considered in this work. The above discussion illustrates that the scenario considered in our work has an array of experimental consequences that can be accessible through multiple avenues. 

Instead of the ${\mathbb Z^\chi_2}$ symmetry, we could have considered $\Phi$ and $X$ to be charged under a U(1) gauge symmetry. In this case other phenomenological opportunities would arise from kinetic mixing terms, such as the possibility of millicharged  matter. We leave a complete phenomenological study to future work.

\section{Summary and Conclusions} 
In this work, we have considered ALL: a dynamical scenario in which low-scale leptogenesis can be realized through a hidden sector, that simultaneously explains the masses of the light neutrinos as well as the relic abundance of dark matter. The scenario rests on the evolution of a scalar field $\phi$, which assumes a large (negative) vacuum expectation value $\vev{\phi} \sim - g_X m_X \xi/H $ when the hidden sector fermion $X$ becomes massive. 
The large $\phi$ values in turn lead to a suddenly and temporarily enhanced Yukawa coupling for a $\sim 10$~TeV sterile neutrino $N_1$, which promptly decays giving rise to a lepton asymmetry.  This asymmetry can be converted to a baryon asymmetry by the electroweak sphalerons. 
After EWSB, there will be a time at which the $\phi$ mass becomes dominant over the Hubble rate, and $\phi$ starts oscillating around $\vev{\phi} \sim -g_X T^2 m_X / (6 m_0^2 + g_X^2 T^2) $, falling with temperature. Then the $N_1$ coupling is also restored to a small value, which we take here to yield a SM neutrino mass much lighter than the other two, possibly vanishing.

We also showed that in our framework, the fermion $X$ can play the role of the dark matter. We demonstrated this in an explicit scenario where the relic abundance is set by $ X \bar{X} \to \Phi \Phi$.  In fact, with a mild departure from the values of parameters assumed in this case, one can also arrive at a scenario where the light scalar $\phi$ can be a significant -- or perhaps a dominant -- component of DM.  Our proposal therefore provides a connection -- which is potentially discernible through multiple experimental signals --  between the processes that produced the visible Universe and the properties of the invisible substance that governs its large scale structure; that is ALL.

\section{Acknowledgements}
The authors thank Lucien Heurtier for useful conversations.  The work of H.D. is supported by the United States Department of Energy under Grant Contract No.~DE-SC0012704. 
D.C. and R.H. are supported by the STFC under Grant No. ST/T001011/1.

\vskip0.5cm

\bibliography{freezein-refs}

\begin{thebibliography}{32}
\expandafter\ifx\csname natexlab\endcsname\relax\def\natexlab#1{#1}\fi
\expandafter\ifx\csname bibnamefont\endcsname\relax
  \def\bibnamefont#1{#1}\fi
\expandafter\ifx\csname bibfnamefont\endcsname\relax
  \def\bibfnamefont#1{#1}\fi
\expandafter\ifx\csname citenamefont\endcsname\relax
  \def\citenamefont#1{#1}\fi
\expandafter\ifx\csname url\endcsname\relax
  \def\url#1{\texttt{#1}}\fi
\expandafter\ifx\csname urlprefix\endcsname\relax\def\urlprefix{URL }\fi
\providecommand{\bibinfo}[2]{#2}
\providecommand{\eprint}[2][]{\url{#2}}

\bibitem[{\citenamefont{Minkowski}(1977)}]{Minkowski:1977sc}
\bibinfo{author}{\bibfnamefont{P.}~\bibnamefont{Minkowski}},
  \bibinfo{journal}{Phys. Lett. B} \textbf{\bibinfo{volume}{67}},
  \bibinfo{pages}{421} (\bibinfo{year}{1977}).

\bibitem[{\citenamefont{Gell-Mann et~al.}(1979)\citenamefont{Gell-Mann, Ramond,
  and Slansky}}]{Gell-Mann:1979vob}
\bibinfo{author}{\bibfnamefont{M.}~\bibnamefont{Gell-Mann}},
  \bibinfo{author}{\bibfnamefont{P.}~\bibnamefont{Ramond}}, \bibnamefont{and}
  \bibinfo{author}{\bibfnamefont{R.}~\bibnamefont{Slansky}},
  \bibinfo{journal}{Conf. Proc. C} \textbf{\bibinfo{volume}{790927}},
  \bibinfo{pages}{315} (\bibinfo{year}{1979}), \eprint{1306.4669}.

\bibitem[{\citenamefont{Mohapatra and Senjanovic}(1980)}]{Mohapatra:1979ia}
\bibinfo{author}{\bibfnamefont{R.~N.} \bibnamefont{Mohapatra}}
  \bibnamefont{and}
  \bibinfo{author}{\bibfnamefont{G.}~\bibnamefont{Senjanovic}},
  \bibinfo{journal}{Phys. Rev. Lett.} \textbf{\bibinfo{volume}{44}},
  \bibinfo{pages}{912} (\bibinfo{year}{1980}).

\bibitem[{\citenamefont{Yanagida}(1979)}]{Yanagida:1979as}
\bibinfo{author}{\bibfnamefont{T.}~\bibnamefont{Yanagida}},
  \bibinfo{journal}{Conf. Proc. C} \textbf{\bibinfo{volume}{7902131}},
  \bibinfo{pages}{95} (\bibinfo{year}{1979}).

\bibitem[{\citenamefont{Schechter and Valle}(1980)}]{Schechter:1980gr}
\bibinfo{author}{\bibfnamefont{J.}~\bibnamefont{Schechter}} \bibnamefont{and}
  \bibinfo{author}{\bibfnamefont{J.~W.~F.} \bibnamefont{Valle}},
  \bibinfo{journal}{Phys. Rev. D} \textbf{\bibinfo{volume}{22}},
  \bibinfo{pages}{2227} (\bibinfo{year}{1980}).

\bibitem[{\citenamefont{Fukugita and Yanagida}(1986)}]{Fukugita:1986hr}
\bibinfo{author}{\bibfnamefont{M.}~\bibnamefont{Fukugita}} \bibnamefont{and}
  \bibinfo{author}{\bibfnamefont{T.}~\bibnamefont{Yanagida}},
  \bibinfo{journal}{Phys. Lett. B} \textbf{\bibinfo{volume}{174}},
  \bibinfo{pages}{45} (\bibinfo{year}{1986}).

\bibitem[{\citenamefont{Zyla et~al.}(2020)}]{ParticleDataGroup:2020ssz}
\bibinfo{author}{\bibfnamefont{P.~A.} \bibnamefont{Zyla}} \bibnamefont{et~al.}
  (\bibinfo{collaboration}{Particle Data Group}), \bibinfo{journal}{PTEP}
  \textbf{\bibinfo{volume}{2020}}, \bibinfo{pages}{083C01}
  (\bibinfo{year}{2020}).

\bibitem[{\citenamefont{Davidson et~al.}(2008)\citenamefont{Davidson, Nardi,
  and Nir}}]{Davidson:2008bu}
\bibinfo{author}{\bibfnamefont{S.}~\bibnamefont{Davidson}},
  \bibinfo{author}{\bibfnamefont{E.}~\bibnamefont{Nardi}}, \bibnamefont{and}
  \bibinfo{author}{\bibfnamefont{Y.}~\bibnamefont{Nir}},
  \bibinfo{journal}{Phys. Rept.} \textbf{\bibinfo{volume}{466}},
  \bibinfo{pages}{105} (\bibinfo{year}{2008}), \eprint{0802.2962}.

\bibitem[{\citenamefont{Manton}(1983)}]{Manton:1983nd}
\bibinfo{author}{\bibfnamefont{N.~S.} \bibnamefont{Manton}},
  \bibinfo{journal}{Phys. Rev. D} \textbf{\bibinfo{volume}{28}},
  \bibinfo{pages}{2019} (\bibinfo{year}{1983}).

\bibitem[{\citenamefont{Klinkhamer and Manton}(1984)}]{Klinkhamer:1984di}
\bibinfo{author}{\bibfnamefont{F.~R.} \bibnamefont{Klinkhamer}}
  \bibnamefont{and} \bibinfo{author}{\bibfnamefont{N.~S.}
  \bibnamefont{Manton}}, \bibinfo{journal}{Phys. Rev. D}
  \textbf{\bibinfo{volume}{30}}, \bibinfo{pages}{2212} (\bibinfo{year}{1984}).

\bibitem[{\citenamefont{Pilaftsis}(1997)}]{Pilaftsis:1997jf}
\bibinfo{author}{\bibfnamefont{A.}~\bibnamefont{Pilaftsis}},
  \bibinfo{journal}{Phys. Rev. D} \textbf{\bibinfo{volume}{56}},
  \bibinfo{pages}{5431} (\bibinfo{year}{1997}), \eprint{hep-ph/9707235}.

\bibitem[{\citenamefont{Pilaftsis and Underwood}(2004)}]{Pilaftsis:2003gt}
\bibinfo{author}{\bibfnamefont{A.}~\bibnamefont{Pilaftsis}} \bibnamefont{and}
  \bibinfo{author}{\bibfnamefont{T.~E.~J.} \bibnamefont{Underwood}},
  \bibinfo{journal}{Nucl. Phys. B} \textbf{\bibinfo{volume}{692}},
  \bibinfo{pages}{303} (\bibinfo{year}{2004}), \eprint{hep-ph/0309342}.

\bibitem[{\citenamefont{Boubekeur et~al.}(2004)\citenamefont{Boubekeur, Hambye,
  and Senjanovic}}]{Boubekeur:2004ez}
\bibinfo{author}{\bibfnamefont{L.}~\bibnamefont{Boubekeur}},
  \bibinfo{author}{\bibfnamefont{T.}~\bibnamefont{Hambye}}, \bibnamefont{and}
  \bibinfo{author}{\bibfnamefont{G.}~\bibnamefont{Senjanovic}},
  \bibinfo{journal}{Phys. Rev. Lett.} \textbf{\bibinfo{volume}{93}},
  \bibinfo{pages}{111601} (\bibinfo{year}{2004}), \eprint{hep-ph/0404038}.

\bibitem[{\citenamefont{Batell and Ghalsasi}(2021)}]{Batell:2021ofv}
\bibinfo{author}{\bibfnamefont{B.}~\bibnamefont{Batell}} \bibnamefont{and}
  \bibinfo{author}{\bibfnamefont{A.}~\bibnamefont{Ghalsasi}}
  (\bibinfo{year}{2021}), \eprint{2109.04476}.

\bibitem[{\citenamefont{Cohen et~al.}(2008)\citenamefont{Cohen, Morrissey, and
  Pierce}}]{Cohen:2008nb}
\bibinfo{author}{\bibfnamefont{T.}~\bibnamefont{Cohen}},
  \bibinfo{author}{\bibfnamefont{D.~E.} \bibnamefont{Morrissey}},
  \bibnamefont{and} \bibinfo{author}{\bibfnamefont{A.}~\bibnamefont{Pierce}},
  \bibinfo{journal}{Phys. Rev. D} \textbf{\bibinfo{volume}{78}},
  \bibinfo{pages}{111701} (\bibinfo{year}{2008}), \eprint{0808.3994}.

\bibitem[{\citenamefont{Baker and Kopp}(2017)}]{Baker:2016xzo}
\bibinfo{author}{\bibfnamefont{M.~J.} \bibnamefont{Baker}} \bibnamefont{and}
  \bibinfo{author}{\bibfnamefont{J.}~\bibnamefont{Kopp}},
  \bibinfo{journal}{Phys. Rev. Lett.} \textbf{\bibinfo{volume}{119}},
  \bibinfo{pages}{061801} (\bibinfo{year}{2017}), \eprint{1608.07578}.

\bibitem[{\citenamefont{Baker and Mittnacht}(2019)}]{Baker:2018vos}
\bibinfo{author}{\bibfnamefont{M.~J.} \bibnamefont{Baker}} \bibnamefont{and}
  \bibinfo{author}{\bibfnamefont{L.}~\bibnamefont{Mittnacht}},
  \bibinfo{journal}{JHEP} \textbf{\bibinfo{volume}{05}}, \bibinfo{pages}{070}
  (\bibinfo{year}{2019}), \eprint{1811.03101}.

\bibitem[{\citenamefont{Davoudiasl and Mohlabeng}(2020)}]{Davoudiasl:2019xeb}
\bibinfo{author}{\bibfnamefont{H.}~\bibnamefont{Davoudiasl}} \bibnamefont{and}
  \bibinfo{author}{\bibfnamefont{G.}~\bibnamefont{Mohlabeng}},
  \bibinfo{journal}{JHEP} \textbf{\bibinfo{volume}{04}}, \bibinfo{pages}{177}
  (\bibinfo{year}{2020}), \eprint{1912.05572}.

\bibitem[{\citenamefont{Croon et~al.}(2022)\citenamefont{Croon, Elor, Houtz,
  Murayama, and White}}]{Croon:2020ntf}
\bibinfo{author}{\bibfnamefont{D.}~\bibnamefont{Croon}},
  \bibinfo{author}{\bibfnamefont{G.}~\bibnamefont{Elor}},
  \bibinfo{author}{\bibfnamefont{R.}~\bibnamefont{Houtz}},
  \bibinfo{author}{\bibfnamefont{H.}~\bibnamefont{Murayama}}, \bibnamefont{and}
  \bibinfo{author}{\bibfnamefont{G.}~\bibnamefont{White}},
  \bibinfo{journal}{Phys. Rev. D} \textbf{\bibinfo{volume}{105}},
  \bibinfo{pages}{L061303} (\bibinfo{year}{2022}), \eprint{2012.15284}.

\bibitem[{\citenamefont{Patt and Wilczek}(2006)}]{Patt:2006fw}
\bibinfo{author}{\bibfnamefont{B.}~\bibnamefont{Patt}} \bibnamefont{and}
  \bibinfo{author}{\bibfnamefont{F.}~\bibnamefont{Wilczek}}
  (\bibinfo{year}{2006}), \eprint{hep-ph/0605188}.

\bibitem[{\citenamefont{Dom\`enech and Sasaki}(2021)}]{Domenech:2021uyx}
\bibinfo{author}{\bibfnamefont{G.}~\bibnamefont{Dom\`enech}} \bibnamefont{and}
  \bibinfo{author}{\bibfnamefont{M.}~\bibnamefont{Sasaki}},
  \bibinfo{journal}{JCAP} \textbf{\bibinfo{volume}{06}}, \bibinfo{pages}{030}
  (\bibinfo{year}{2021}), \eprint{2104.05271}.

\bibitem[{\citenamefont{Quiros}(1999)}]{Quiros:1999jp}
\bibinfo{author}{\bibfnamefont{M.}~\bibnamefont{Quiros}}, in
  \emph{\bibinfo{booktitle}{{ICTP Summer School in High-Energy Physics and
  Cosmology}}} (\bibinfo{year}{1999}), pp. \bibinfo{pages}{187--259},
  \eprint{hep-ph/9901312}.

\bibitem[{\citenamefont{Davoudiasl et~al.}(2016)\citenamefont{Davoudiasl,
  Hooper, and McDermott}}]{Davoudiasl:2015vba}
\bibinfo{author}{\bibfnamefont{H.}~\bibnamefont{Davoudiasl}},
  \bibinfo{author}{\bibfnamefont{D.}~\bibnamefont{Hooper}}, \bibnamefont{and}
  \bibinfo{author}{\bibfnamefont{S.~D.} \bibnamefont{McDermott}},
  \bibinfo{journal}{Phys. Rev. Lett.} \textbf{\bibinfo{volume}{116}},
  \bibinfo{pages}{031303} (\bibinfo{year}{2016}), \eprint{1507.08660}.

\bibitem[{\citenamefont{Kaplinghat et~al.}(2014)\citenamefont{Kaplinghat,
  Tulin, and Yu}}]{Kaplinghat:2013yxa}
\bibinfo{author}{\bibfnamefont{M.}~\bibnamefont{Kaplinghat}},
  \bibinfo{author}{\bibfnamefont{S.}~\bibnamefont{Tulin}}, \bibnamefont{and}
  \bibinfo{author}{\bibfnamefont{H.-B.} \bibnamefont{Yu}},
  \bibinfo{journal}{Phys. Rev. D} \textbf{\bibinfo{volume}{89}},
  \bibinfo{pages}{035009} (\bibinfo{year}{2014}), \eprint{1310.7945}.

\bibitem[{\citenamefont{Aprile et~al.}(2018)}]{XENON:2018voc}
\bibinfo{author}{\bibfnamefont{E.}~\bibnamefont{Aprile}} \bibnamefont{et~al.}
  (\bibinfo{collaboration}{XENON}), \bibinfo{journal}{Phys. Rev. Lett.}
  \textbf{\bibinfo{volume}{121}}, \bibinfo{pages}{111302}
  (\bibinfo{year}{2018}), \eprint{1805.12562}.

\bibitem[{\citenamefont{Andersen
  et~al.}(2013)}]{LHCHiggsCrossSectionWorkingGroup:2013rie}
\bibinfo{author}{\bibfnamefont{J.~R.} \bibnamefont{Andersen}}
  \bibnamefont{et~al.} (\bibinfo{collaboration}{LHC Higgs Cross Section Working
  Group}) (\bibinfo{year}{2013}), \eprint{1307.1347}.

\bibitem[{\citenamefont{Aaboud et~al.}(2018)}]{ATLAS:2018pvw}
\bibinfo{author}{\bibfnamefont{M.}~\bibnamefont{Aaboud}} \bibnamefont{et~al.}
  (\bibinfo{collaboration}{ATLAS}), \bibinfo{journal}{JHEP}
  \textbf{\bibinfo{volume}{10}}, \bibinfo{pages}{031} (\bibinfo{year}{2018}),
  \eprint{1806.07355}.

\bibitem[{\citenamefont{Tumasyan et~al.}(2022)}]{CMS:2022qva}
\bibinfo{author}{\bibfnamefont{A.}~\bibnamefont{Tumasyan}} \bibnamefont{et~al.}
  (\bibinfo{collaboration}{CMS}) (\bibinfo{year}{2022}), \eprint{2201.11585}.

\bibitem[{\citenamefont{Aad et~al.}(2022)}]{ATLAS:2022yvh}
\bibinfo{author}{\bibfnamefont{G.}~\bibnamefont{Aad}} \bibnamefont{et~al.}
  (\bibinfo{collaboration}{ATLAS}) (\bibinfo{year}{2022}), \eprint{2202.07953}.

\bibitem[{\citenamefont{Knapen et~al.}(2017)\citenamefont{Knapen, Lin, and
  Zurek}}]{Knapen:2017xzo}
\bibinfo{author}{\bibfnamefont{S.}~\bibnamefont{Knapen}},
  \bibinfo{author}{\bibfnamefont{T.}~\bibnamefont{Lin}}, \bibnamefont{and}
  \bibinfo{author}{\bibfnamefont{K.~M.} \bibnamefont{Zurek}},
  \bibinfo{journal}{Phys. Rev. D} \textbf{\bibinfo{volume}{96}},
  \bibinfo{pages}{115021} (\bibinfo{year}{2017}), \eprint{1709.07882}.

\bibitem[{\citenamefont{Adelberger et~al.}(2009)\citenamefont{Adelberger,
  Gundlach, Heckel, Hoedl, and Schlamminger}}]{Adelberger:2009zz}
\bibinfo{author}{\bibfnamefont{E.~G.} \bibnamefont{Adelberger}},
  \bibinfo{author}{\bibfnamefont{J.~H.} \bibnamefont{Gundlach}},
  \bibinfo{author}{\bibfnamefont{B.~R.} \bibnamefont{Heckel}},
  \bibinfo{author}{\bibfnamefont{S.}~\bibnamefont{Hoedl}}, \bibnamefont{and}
  \bibinfo{author}{\bibfnamefont{S.}~\bibnamefont{Schlamminger}},
  \bibinfo{journal}{Prog. Part. Nucl. Phys.} \textbf{\bibinfo{volume}{62}},
  \bibinfo{pages}{102} (\bibinfo{year}{2009}).

\bibitem[{\citenamefont{Heeck}(2014)}]{Heeck:2014zfa}
\bibinfo{author}{\bibfnamefont{J.}~\bibnamefont{Heeck}},
  \bibinfo{journal}{Phys. Lett. B} \textbf{\bibinfo{volume}{739}},
  \bibinfo{pages}{256} (\bibinfo{year}{2014}), \eprint{1408.6845}.

\end{thebibliography}

\end{document}